\documentclass[prc,aps,preprint,showkeys,showpacs,nofootinbib]{revtex4}
\usepackage{graphicx}

\begin{document}

\date{\today}
\title{\textbf{Optimizing stochastic trajectories in exact quantum jump
approaches of interacting systems}}
\author{Denis Lacroix}

\address{Laboratoire de Physique Corpusculaire, \\
ENSICAEN and Universit\'e de Caen,IN2P3-CNRS,\\
Blvd du Mar\'{e}chal Juin
14050 Caen, France}
\begin{abstract}
The quantum jump approach, where pairs of state vectors follow Stochastic
Schroedinger Equation (SSE) in order to treat the exact quantum dynamics of two
interacting systems, is first described. 
In this work the non-uniqueness of such stochastic Schroedinger equations  
is investigated to propose strategies to optimize the stochastic paths 
and reduce statistical fluctuations. 
In the proposed method, called the 
'\textit{adaptative noise method}', a specific SSE is obtained for which the
noise depends explicitly on both the initial state and on the properties of
the interaction Hamiltonian. It is also shown that this method can be
further improved by introduction of a mean-field dynamics. The
different optimization procedures are illustrated quantitatively in the case
of interacting spins. A significant reduction of the statistical
fluctuations is obtained. Consequently a much smaller number of
trajectories is needed to accurately reproduce the exact dynamics as
compared to the SSE without optimization.
\end{abstract}


\maketitle

Recently the description of open quantum systems with Stochastic Schroedinger Equation (SSE) 
has received much attention 
\cite{Ple98,Bre02,Sto02}. This is due firstly to the possibility of substituting the
description of a complex system based on the evolution of its density matrix
by a description based on the stochastic evolution of wave-packets. 
Therefore these methods reduce 
significantly the number of degrees of freedom to be taken into account. A
second reason is the increasing computational facilities that allow for massive
numerical applications. The Monte-Carlo wave-function techniques are
presently  extensively used to treat Markovian master equations in the
Lindblad form \cite{Dal92,Dum92,Gis92,Car93,Cas96,Ple98,Ima94,Gar00,Bre02}. 
In that case the system density evolution (noted $\rho_S$) 
is simulated using SSE on system state-vectors 
$\left| \Phi \right>$. Then the dissipative evolution is recovered by averaging 
symmetric densities $\rho = \left| \Phi  \right>\left< \Phi  \right|$ over
different stochastic trajectories, 
i.e. $\rho_S = \overline{\rho}$. Following this strategy, 
various quantum jump approaches based on SSE have been 
proposed \cite{Rig96,Ple98,Bre02}. 
 
Large theoretical efforts have been also devoted to the introduction of 
non-Markovian effects using quantum jumps. A possible way to treat this problem is to use 
stochastic equations that contain a non-local memory kernel \cite{Dio96,Dio98,Str99}. 
In that case the average evolution of the density matrix is still given by averaging over
symmetric densities $\rho=\left| \Phi  \right>\left< \Phi  \right|$, but state vectors 
evolve 
according to integro-differential stochastic equations. 
An alternative method that
avoids the evaluation of non-local memory kernels is to introduce pairs of state-vectors leading 
to densities of the type
$\rho=\left| \Psi_1 \right>\left< \Psi_2 \right|$. Then it is possible 
to treat non-Markovian effects approximately \cite{Bre99} using time-local SSE. 
More recently \cite{Bre04}
SSE using pairs of state vectors have been proposed as a way to simulate  the 
dynamics of interacting systems exactly, with the difference that $\left| \Psi_1 \right>$ and 
$\left| \Psi_2 \right>$ are separable states of the system and its environment. 
Such an exact reformulation has already been applied to
different model cases \cite{Bre04} and appears as a candidate to treat the
correlated dynamics of self-interacting mesoscopic systems \cite{Car01,Jul02}
and of systems interacting with an environment \cite{Sha04}.

A common  aspect to all stochastic methods used to simulate 
Markovian and non-Markovian dynamics is that the equation of motion of the density 
matrix does not define uniquely the stochastic equation of state vectors
(see discussion in ref. \cite{Rig96}). 
A specific choice of equation is generally 
retained by imposing additional conditions along the stochastic path. 
For instance in the Markovian limit, nonlinear state-dependent 
stochastic equations have been obtained 
by invoking either normalization conditions of state-vectors, 
orthogonal-jump processes \cite{Dio85,Dio94} or measurement arguments \cite{Gis84,Bre95}.
In addition the introduction of mean-field dynamics \cite{Car01,Jul02} significantly improves 
 the stochastic description.  
Further freedom exists due to the fact that the noise itself can be multiplied 
by an arbitrary phase-factor without breaking the stochastic 
reformulation. Again, this fact has been noted in different studies on Markovian dynamics \cite{Rig96} 
and is sometimes used to pass from one stochastic equation to another \cite{Bre02,Ghi90}. In the 
non-Markovian case it can also 
provide appropriate modifications of the stochastic dynamics\cite{Bas03}.

Although an infinite number of stochastic formulations can generally describe the dynamics 
of an open system, the physical interpretation as well as the nature of the process 
can be completely different from one SSE to another. In addition, a series of works dedicated 
to Markovian dynamics with symmetric densities have 
pointed out that stochastic formulations are not equivalent 
as far as numerical implementation is concerned \cite{Gil97,Gar00,Car01}.
For instance numerical efficiency 
can be significantly improved by considering evolution of normalized states. 
The flexibility of SSE approaches has however not been explored yet when pairs of states vectors 
are used. 
It is clear that, in order to be able to apply the latter theory
to a large variety of physical problems, specific accurate methods 
must be developed.
 
The main purpose of this paper is to investigate more systematically the 
freedom in the formulation of stochastic process using pairs of states in 
order to  optimize 
the quantum jumps and reduce the
number of trajectories necessary to describe  the dynamics of
interacting systems accurately. 
The paper is organised as follows: first the
procedure to reinterpret the dynamics of interacting systems in terms of a
Monte-Carlo evolution of pairs of wave-packets is described. The existence of an infinite 
number of SSE dynamics leading to
a large freedom in defining stochastic trajectories is illustrated. 
In section \ref{secopt} general strategies using this freedom to optimize the
quantum jumps and minimize the statistical fluctuations are described. In this case we
show that the optimal noise depends explicitly on the initial state and on
the properties of the interaction. In section \ref{secsmf} we show that the
combination of mean-field dynamics with the optimization of the noise leads
to an additional reduction of the statistical fluctuations. Lastly, the method
is illustrated quantitatively with a schematic model of interacting spin
systems.

\section{SSE for the exact dynamics of interacting systems}

Here we give a guideline for the exact reformulation of correlated dynamics in
terms of diffusive wave-function processes using pairs of wave-packets. 
Note that a formal derivation of the equation 
given below can be obtained using 
the Hubbard-Stratonovich transformation \cite{Hub59,Str58}, see for instance ref. \cite{Sha04}.

\subsection{SSE with pairs of wave-packets}
\label{gen_dif}

Following \cite{Bre04} we start
from a general Hamiltonian describing the interaction of a system and its
environment given by: 
\begin{equation}
H_{I}=\sum_{\alpha }A_{\alpha } \otimes B_{\alpha }  \label{eq:ham}
\end{equation}
Here we assume that the Hamiltonian is possibly already written in the
interaction picture. $A_{\alpha }$ and $B_{\alpha }$ are operators acting
respectively on the system and on the environment. Starting from an initial
uncorrelated state: 
\begin{eqnarray}
\left| \Psi \right>= \left| \Phi \right\rangle \otimes \left| \chi
\right\rangle,  \label{eq:indep}
\end{eqnarray}
where $\left| \Phi \right\rangle$ and $\left| \chi \right\rangle$ are state
vectors for the system and the environment respectively, the evolution of
the system can be replaced by the set of stochastic Schroedinger equations: 
\begin{equation}
\left\{ 
\begin{array}{l}
d\left| \Phi\right\rangle = \gamma \sum_{\alpha }a_\alpha \left( t\right)
A_{\alpha } \left| \Phi \right\rangle \\ 
d\left| \chi \right\rangle = \gamma \sum_{\alpha }b_\alpha \left( t\right)
B_{\alpha } \left| \chi \right\rangle%
\end{array}
\right.  \label{eq:sse}
\end{equation}
where $\gamma $ is a parameter defined formally as $\gamma =\sqrt{\frac{d t}{%
i\hbar }}$. Here $a_\alpha(t)$ and $b_\alpha(t)$ are complex stochastic
Gaussian variables that follow the Ito stochastic rules \cite{Gar85}. Note
that although we consider diffusive Gaussian processes, other stochastic
methods could be used like piecewise deterministic processes (PDP's)
\cite{Dio86,Bre04}. Indeed, the results presented in this work can be adapted to PDP
quantum mechanics. In the following, we note $\overline {X}$ the average of $%
X$ over the stochastic variables. Under the condition: 
\begin{eqnarray}
\overline{a_\alpha(t) b_\alpha(t)}= 1,  \label{condab}
\end{eqnarray}
the average evolution corresponds to the exact Schroedinger equation 
\begin{eqnarray}
\overline{d\left| \Psi \right>} =\frac{ d t }{i\hbar} H_I\left| \Psi \right>.
\end{eqnarray}
An interesting aspect of this random process is that stochastic equations 
(\ref{eq:sse}) preserve the separability of the total state given by eq. (\ref%
{eq:indep}) so that the procedure can be iterated to simulate the evolution
on large time scales.

A second attractive aspect is that the stochastic formulation can be
extended to provide an exact treatment of the Liouville-von Neumann equation
of the density matrix. In that case, we consider
\begin{eqnarray}
D=\left| \Psi_1 \right>\left< \Psi_2 \right|  \label{eq:DDD}
\end{eqnarray}
where $\left| \Psi_1 \right>$ and $\left| \Psi_2 \right>$ is a pair 
of different system+environment states both given by eq. (\ref{eq:indep}). Both states
follow independent stochastic equations given (for both) by eq. (\ref{eq:sse}%
). Noting $\rho(t) = \overline{D(t)} $, its evolution reads 
\begin{eqnarray}
i\hbar \frac{ d \rho }{dt }=\left[H_I , \rho \right]
\end{eqnarray}
which is nothing but the exact Liouville-von Neumann equation. Again, eq. (%
\ref{eq:DDD}) is preserved along each stochastic path. The latter
reformulation of exact dynamics is not limited to initially uncorrelated
states. Indeed if the initial density matrix is correlated, it can be
replaced by an ensemble of couples of states. The initial density reads: $%
\rho(t=0) = E\left( \left| \Psi_1 \right>\left< \Psi_2 \right| \right)$
where the average here means the average over initial dyadics. Then the
complete dynamics is obtained by averaging both on the initial ensemble and
on the stochastic paths.

The random process presented in this section describes
the exact dynamics of a system coupled to an environment. Therefore, although 
the SSE on wave-functions is Markovian, it contains all non-Markovian effects. 
It is also worth noticing that this method significantly differs  from
quantum jump processes used in the Markovian limit 
\cite{Dal92,Dum92,Gis92,Car93,Cas96,Ple98,Ima94,Gar00,Bre02} 
or in the non-Markovian limit with a non-local memory kernel 
\cite{Dio96,Dio98,Str99}. The first reason is that 
pairs of state vectors that evolve according to independent
SSE should be considered. 
A second important aspect is that quantum jumps applied to Markovian 
dynamics require in general to follow only states of the system. Indeed 
the effect of the environment has already been approximated in the Lindblad 
equation describing the system density matrix evolution. 
Here wave-packets of both system and environment should be followed in time and 
the system evolution can be obtained using the $\rho_S = Tr_E(\rho)$ where $Tr_E(.)$ 
denotes the partial trace over the environment. However the necessity 
to follow environment degrees of freedom may lead to additional difficulties.    

\subsection{Statistical Fluctuations}

The possibility to perform exact dynamics of interacting systems is of
particular interest to discuss dissipative effects due to the coupling
of a system with its environment. In particular it does not have the
limitations of Lindblad Master equations \cite{Gar00,Bre04}. 
However large numerical efforts are needed to treat exactly a physical process 
due to the number of trajectories
required to reduce statistical fluctuations of the observables. This is why
so far it has essentially been applied to rather schematic models.

Following \cite{Car01}, a measure of the increase of statistical
fluctuations is given by: 
\begin{equation}
\begin{array}{ll}
\Lambda _{stat} & =E\left( \overline{Tr\left( \left| \left| D\left( t\right)
-\overline{D\left( t\right) }\right| \right|^2 \right)} \right) \\ 
& =E\left( \overline{Tr\left( D^{+}\left( t\right) D\left( t\right) \right)}%
\right) -Tr\left( \rho ^{2}\left( t\right) \right)%
\end{array}
\label{eq:stat}
\end{equation}
where $\left| \left| A \right| \right|^2 = Tr \left( A^+A \right)$ is the
Hilbert-Schmidt norm. In addition, if the initial state is a pure state, 
$Tr\left( \rho ^{2}\left( t\right) \right) =1$.

The growth of statistical fluctuations is of particular importance for
numerical implementations since it is directly connected to the number of
trajectories required to properly reproduce the evolution of a physical
system using SSE. 
Now let us make explicit the evolution of the
statistical noise associated to the SSE defined by eq. (\ref{eq:sse}). We
consider the contribution of a single initial couple of state vectors $%
D=\left| \Psi_1 \right>\left< \Psi_2 \right|$ and we note $\lambda_{stat} =%
\overline{Tr\left( D^{+}\left( t\right) D\left( t\right) \right)}$ its
contribution to $\Lambda_{stat}$. It is assumed for simplicity that each
wave-packet is initially normalized. Starting from $D$, the infinitesimal increase of $%
\lambda_{stat}$ is given by: 
\begin{equation}
\begin{array}{ll}
d\lambda _{stat} & = \overline{ d\left< \Psi_1\left| \right. \Psi_1 \right>}%
+ \overline{ d\left< \Psi_2 \left| \right. \Psi_2 \right>}%
\end{array}%
\end{equation}
in which each contribution reads: 
\begin{eqnarray}
\overline{d\left<\Psi_i \left| \right. \Psi_i \right>} &=& \frac{dt}{\hbar }%
\sum_{\alpha } \left( \overline{ \left| a_\alpha(t) \right|^2 } \left\langle
A_{\alpha }^{+}A_{\alpha }\right\rangle _{\Phi_i } +\overline{ \left|
b_\alpha(t) \right|^2 } \left\langle B_{\alpha }^{+}B_{\alpha }\right\rangle
_{\chi_i }\right.  \nonumber \\
&& \left.+2\Re \left\{ \overline{ a_\alpha(t) b^*_\alpha(t)} \left< A_\alpha
\right>_{\Phi_i }\left< B^+_\alpha \right>_{\chi_i }\right\} \right)
\label{eq:norm1}
\end{eqnarray}
with $i=1,2$. 
Therefore statistical fluctuations depend explicitly on
the properties of the random variables $a_\alpha$ and $b_\alpha $. In
several works \cite{Car01,Jul02,Bre04} specific choices of noise have
been made. In all cases it was possible to demonstrate that the
fluctuations have an upper bound that grows exponentially in
time. Conjointly, the numerical implementation depends 
strongly on the retained SSE as well as the possible 
appearance of instabilities in calculations\cite{Gil97,Gar00}. 
Let us consider more systematically 
the dependence of statistical errors with respect to the freedom
in the definition of SSE using pairs of state vectors.
In the next section we  
specify and use the flexibility in the 
definition of SSE given by equation (\ref{eq:sse}) to show that an
optimal choice of the noise can reduce statistical fluctuations.

\section{Optimal quantum jumps and the reduction of statistical fluctuations}

\label{secopt}

Let us now detail the properties of invariance of the SSE approaches.
Starting from a general complex noise $a_{\alpha }$ and $b_{\alpha }$ that
fulfills the necessary condition given by (\ref{condab}), any new couple of
random variables $\left( a_{\alpha }^{\prime }(t),b_{\alpha }^{\prime
}(t)\right) $ defined by means of the transformation 
\begin{eqnarray}
\left\{ 
\begin{array}{l}
a_{\alpha }^{\prime }(t)=c_{\alpha }a_{\alpha }(t) \\ 
b_{\alpha }^{\prime }(t)=\frac{1}{c_{\alpha }}b_{\alpha }(t)\label{scale1}%
\end{array}%
\right. 
\end{eqnarray}%
with 
\begin{eqnarray}
c_{\alpha }=e^{i\theta _{\alpha }}\sqrt{u_{\alpha }}
\end{eqnarray}%
also gives the correct exact dynamics. In the following, $u_{\alpha }$ and $%
\theta _{\alpha }$ will be referred to respectively the scaling factor and
the phase factor. 
Invariance of the stochastic reformulation with respect to application of a scaling or a 
phase shows that an infinite number of SSE's exists to simulate the exact dynamics of interacting systems, 
as it has been shown in most of the stochastic theories in Hilbert Space \cite{Rig96,Bre02,Bas03}. 
Such an invariance and in particular the invariance with respect
to a phase factor has already been noted in several works \cite{Bre02,Bas03} and
has been used for different purposes. 

Let us now apply a scaling and a phase factor on the 
statistical errors.
We see that, while statistical fluctuations associated to the initial stochastic
equations are given by eq. (\ref{eq:norm1}), the new SSE leads to different
statistical fluctuations given by 
\begin{eqnarray}
\begin{array}{lll}
\overline{d\left\langle \Psi \left\vert {}\right. \Psi \right\rangle }
=\frac{dt}{\hbar }\sum_{\alpha } &&\left( u_{\alpha }\overline{\left\vert a_{\alpha
}(t)\right\vert ^{2}}\left\langle A_{\alpha }^{+}A_{\alpha }\right\rangle
_{\Phi }+\frac{1}{u_{\alpha }}\overline{\left\vert b_{\alpha }(t)\right\vert
^{2}}\left\langle B_{\alpha }^{+}B_{\alpha }\right\rangle _{\chi }\right.  \\
&&\left. +2\Re \left\{ e^{i2\theta _{\alpha }}\overline{a_{\alpha
}(t)b_{\alpha }^{\ast }(t)}\left\langle A_{\alpha }\right\rangle _{\Phi
}\left\langle B_{\alpha }^{+}\right\rangle _{\chi }\right\} \right) 
\end{array}
\label{dff}
\end{eqnarray}%
where the indices $i=1,2$ have been omitted for simplicity. Thus, the growth
of statistical fluctuations may significantly differ depending on the noise. 
In this section we show that the scaling factor as well as the
phase factor can be properly adjusted to obtain minimal statistical
fluctuations thus reducing the number of stochastic trajectories in 
numerical implementations. It may be
seen in the last equation that the parameters of the transformation act
independently on the two parts of eq. (\ref{dff}) and as such, they can be
adjusted separately. For a component $\alpha $ of the interaction
Hamiltonian, the two functions $\Omega _{\alpha }$ and $\Gamma
_{\alpha }$ are defined as: 
\begin{eqnarray}
\Omega _{\alpha }\left( u_{\alpha }\right)  &=&u_{\alpha }\overline{%
\left\vert a_{\alpha }(t)\right\vert ^{2}}\left\langle A_{\alpha
}^{+}A_{\alpha }\right\rangle _{\Phi }+\frac{1}{u_{\alpha }}\overline{%
\left\vert b_{\alpha }(t)\right\vert ^{2}}\left\langle B_{\alpha
}^{+}B_{\alpha }\right\rangle _{\chi }  \label{eq:om} \\
\Gamma _{\alpha }(\theta _{\alpha }) &=&2\Re \left\{ e^{i2\theta _{\alpha }}%
\overline{a_{\alpha }(t)b_{\alpha }^{\ast }(t)}\left\langle A_{\alpha
}\right\rangle _{\Phi }\left\langle B_{\alpha }^{+}\right\rangle _{\chi
}\right\}   \label{eq:gam}
\end{eqnarray}%
Starting from a given statistical noise $\left( a_{\alpha }(t),b_{\alpha
}(t)\right) $, the strategy is to find the optimal $u_{\alpha }$ and $\theta
_{\alpha }$ that minimize respectively these two functions.

\subsection{Optimal scaling factor}

In eq. (\ref{eq:om}) both $\left\langle A_{\alpha }^{+}A_{\alpha
}\right\rangle _{\Phi }$ and $\left\langle B_{\alpha }^{+}B_{\alpha
}\right\rangle _{\chi }$ are fixed parameters that depend on the initial
state. For a given initial state vector, $\Omega _\alpha$ is minimal for 
\begin{eqnarray}
u_\alpha = \left(\frac{ \overline{\left| b_\alpha \right|^2} }{ \overline{%
\left| a_\alpha \right|^2}} \frac{\left\langle B_{\alpha }^{+}B_{\alpha
}\right\rangle _{\chi }} {\left\langle A_{\alpha }^{+}A_{\alpha
}\right\rangle _{\Phi } }\right)^{1/2}  \label{eq:sfact}
\end{eqnarray}
Under this specific choice the lower limit of $\Omega_\alpha $ reads: 
\begin{eqnarray}
\Omega _\alpha=2 \sqrt{\overline{\left| a_\alpha(t) \right|^2} \cdot 
\overline{\left| b_\alpha(t) \right|^2} \left\langle A_{\alpha
}^{+}A_{\alpha }\right\rangle _{\Phi } \left\langle B_{\alpha}^{+}B_{\alpha
}\right\rangle _{\chi } }
\end{eqnarray}
This expression is valid for any $a_\alpha $ and $b_\alpha $ fulfilling the condition (%
\ref{condab}). Due to eq. (\ref{condab}), we also have $\overline{%
\left| a_\alpha(t) \right|^2} \cdot \overline{\left| b_\alpha(t) \right|^2}
\geq 1$. Thus it is better to start with a noise such that $\overline{%
\left| a_\alpha(t) \right|^2} \cdot \overline{\left| b_\alpha(t) \right|^2}
= 1$. Coming back to the general case, it is important to note that the
optimization depends on the state on which the expectation values are taken.
In particular, if 
\begin{eqnarray}
\left( \frac{ \overline{\left| b_\alpha \right|^2} }{ \overline{\left|
a_\alpha \right|^2}}\frac{\left\langle B_{\alpha }^{+}B_{\alpha
}\right\rangle _{\chi }} {\left\langle A_{\alpha }^{+}A_{\alpha
}\right\rangle _{\Phi } }\right)=1,
\end{eqnarray}
the minimal value of $\Omega _\alpha $ corresponds exactly to expression (%
\ref{dff}). In this case, no reduction of the statistical fluctuation is
obtained. Such a case will be presented in a forthcoming section. 
Generally speaking, the more the optimal value of $u_\alpha $ differs from one,
the larger the reduction of the statistical noise.

\subsection{Optimal phase factor}

\label{optph}

Let us now turn to the second term in eq. (\ref{dff}). In order to minimize
the statistical noise, the second term should be negative. It is always
possible to fix conveniently $\theta_\alpha$ in order to have the minimum
value for $\Gamma _\alpha $, i.e: 
\begin{eqnarray}
\Gamma _\alpha = - 2 \left|\overline{ a_\alpha b^*_\alpha}\right| \cdot \left|
\left< A_\alpha \right>_{\Phi } \left< B^+_\alpha \right>_{\chi } \right|
\label{dff_caseopt}
\end{eqnarray}
The absolute value of the expression depends explicitly on $\left|%
\overline{a_\alpha b^*_\alpha} \right|$. Let us now give an explicit
expression for $a_\alpha$ and $b_\alpha$. We define 
\begin{eqnarray}
\left\{ 
\begin{array}{l}
a_\alpha = x_\alpha +i y_\alpha \\ 
b_\alpha = x^{\prime}_\alpha +i y^{\prime}_\alpha \label{xy}%
\end{array}
\right.
\end{eqnarray}
where all components are real Gaussian stochastic variables. Using condition
(\ref{condab}), we obtain: 
\begin{eqnarray}
\left| \overline{ a_\alpha b^*_\alpha }\right|^2 &=& -1 + 2\left( \overline{%
x_\alpha x^{\prime}_\alpha}^2 +\overline{ y_\alpha y^{\prime}_\alpha }^2 + 
\overline{ y_\alpha x^{\prime}_\alpha}^2 + \overline{x_\alpha
y^{\prime}_\alpha }^2 \right)  \label{absab}
\end{eqnarray}
which leads to the inequality 
\begin{eqnarray}
\left| \overline{a_\alpha b^*_\alpha} \right|^2 \leq -1 + 2 \overline{
\left|a_\alpha^2 \right|} \cdot \overline{ \left|b_\alpha^2 \right|}.
\end{eqnarray}
In order to reduce the fluctuations, expression (\ref{absab}) must be
maximized. The maximal value is obtained if $a_\alpha$ and $b^*_\alpha$ are fully correlated, 
i.e. $b^*_\alpha \propto a_\alpha$. 
Therefore a convenient choice for the couple $(a_{\alpha },b_{\alpha })$ is 
\begin{eqnarray}
\begin{array}{lll}
a_{\alpha } & = & \sqrt{\delta }e^{i\varphi }x_{\alpha } \\ 
b_{\alpha } & = & \frac{1}{\sqrt{\delta }}e^{-i\varphi }x_{\alpha }%
\end{array}%
\label{eq:aabb}
\end{eqnarray}%
where $\overline{x_{\alpha }^{2}}=1$ and $\delta$ is a real parameter. 
The latter expression leads to $\left\vert \overline{a_{\alpha }b_{\alpha }^{\ast }}%
\right\vert ^{2}=-1+2\overline{\left\vert a_{\alpha }^{2}\right\vert }\cdot 
\overline{\left\vert b_{\alpha }^{2}\right\vert }=1$. Since any noise given
by the expression (\ref{eq:aabb}) can be obtained from an initial real noise $x_{\alpha }$
through the transformation (\ref{scale1}), we simply assume 
that $a_{\alpha }=b_{\alpha }=x_{\alpha }$. Then,
if we define $\theta _{AB}$ as: 
\begin{eqnarray}
\left\langle A_{\alpha }\right\rangle _{\Phi }\left\langle B_{\alpha
}^{+}\right\rangle _{\chi }=e^{i\theta _{AB}}\left\vert \left\langle
A_{\alpha }\right\rangle _{\Phi }\left\langle B_{\alpha }^{+}\right\rangle
_{\chi }\right\vert 
\end{eqnarray}%
the optimal phase factor given by 
\begin{eqnarray}
2\theta _{\alpha }=\pi -\theta _{AB}
\label{eq:tfact}
\end{eqnarray}%
leads to the minimum value
\begin{eqnarray}
\Gamma _{\alpha }=-2\left\vert \left\langle A_{\alpha }\right\rangle _{\Phi
}\left\langle B_{\alpha }^{+}\right\rangle _{\chi }\right\vert .
\end{eqnarray}

\subsection{Summary}

In this section starting from the diffusive equation given in section \ref{gen_dif}, we have
shown that the noise can be optimized to reduce the statistical
fluctuations. For any given initial state $\left| \Psi \right>= \left| \Phi
\right\rangle \otimes \left| \chi \right\rangle$, an optimal reduction of
the statistical fluctuations corresponds to a diffusive process with a
specific complex Gaussian noise given by 
\begin{eqnarray}
\left\{ 
\begin{array}{l}
a_\alpha (t)= e^{i\theta_\alpha} \sqrt{u_\alpha} x_\alpha \\ 
b_\alpha (t)= e^{-i\theta_\alpha} \frac{ 1 }{\sqrt{u_\alpha} } x_\alpha %
\label{scale}%
\end{array}
\right.
\end{eqnarray}
with $\overline{x^2_\alpha}=1$. In these expressions both the scaling factor
and the phase factor, given respectively by equations (\ref{eq:sfact}) and 
(\ref{eq:tfact}) depend explicitly on the initial state vector as well as on
the Hamiltonian. Along the optimal path, the normalization evolution reads: 
\begin{eqnarray}
\overline{d\left<\Psi \left| \right. \Psi \right>} &= & 2\frac{d t}{\hbar }%
\sum_{\alpha } \left( \sqrt{ \left\langle A_{\alpha }^{+}A_{\alpha
}\right\rangle _{\Phi } \left\langle B_{\alpha}^{+}B_{\alpha }\right\rangle
_{\chi } } - \sqrt{\left< A^+_\alpha \right>_{\Phi }\left< A_\alpha
\right>_{\Phi }} \sqrt{\left< B^+_\alpha \right>_{\chi }\left< B_\alpha
\right>_{\chi }} \right) \\
&\equiv& 2\frac{d t}{\hbar }\sum_{\alpha } F_\alpha  \label{dff_opt}
\end{eqnarray}
The procedure described here for a single initial state can be applied to
the complete description of correlated systems. In this case, an ensemble of
wave-packets is considered simultaneously, each state evolving according to
its own stochastic equation with a noise adapted at each time-step. Since
for all states the trajectories are optimally chosen, we do expect that the
total statistical fluctuations obtained by averaging over the ensemble of
states will also be reduced. In the following, this method will be referred
to as the 'adaptative noise method'.

\section{Noise on top of mean-field dynamics}

\label{secsmf}

Before giving an example of application of the adaptative noise procedure,
we would like to mention that the flexibility on the noise is not the only
way to optimize quantum jumps.  
It is possible to introduce a
deterministic part in addition to the stochastic contribution that partially treats
  the coupling of the system with the environment without breaking
the separability of the state vector. 
{ One ends up with non-linear, state dependent, stochastic 
equations. Such non-linear equations have been obtained in the Markovian limit
by assuming jumps orthogonal to the state \cite{Dio85,Dio86,Dio94} or "optimal" 
measurement arguments \cite{Gis84,Bre95}. In all cases we expect that the 
numerical accuracy will be improved. Closely related is the introduction of 
mean-field dynamics associated to the interacting system  \cite{Car01,Jul02,Sha04}.    

In this section the latter method is used.   
The mean-field associated to the general Hamiltonian given by 
(\ref{eq:ham}) as well as the associated stochastic dynamics are described. 
As will be seen, the introduction
of mean-field gives non-linear equations equivalent to other techniques. 
Lastly we discuss how the attractive aspects inherent to mean-field theories and to
the adaptative noise can be combined to take advantage of both methods.
}

\subsection{Optimized Stochastic mean-field dynamics}

Starting from the interaction Hamiltonian the associated mean-field
dynamics can be obtained using the variational principle 
\begin{eqnarray}
\delta \left\langle \Psi \left\vert i\hbar \partial _{t}-H_{I}\right\vert
\Psi \right\rangle =0
\end{eqnarray}%
where $\left\vert \Psi \right\rangle $ is a separable state given by
equation (\ref{eq:indep}). Using the variational principle we obtain the
mean-field equation of motion for each component of the total state vector. 
\begin{equation}
\left\{ 
\begin{array}{lll}
i\hbar \partial _{t}\left\vert \Phi \right\rangle = & \sum_{\alpha
}\left\langle B_{\alpha }\right\rangle _{\chi }A_{\alpha }\left\vert \Phi
\right\rangle = & h_{MF}^{S}\left\vert \Phi \right\rangle  \\ 
&  &  \\ 
i\hbar \partial _{t}\left\vert \chi \right\rangle = & \sum_{\alpha
}\left\langle A_{\alpha }\right\rangle _{\Phi }B_{\alpha }\left\vert \chi
\right\rangle = & h_{MF}^{E}\left\vert \chi \right\rangle 
\end{array}%
\right. 
\end{equation}%
where $h_{MF}^{S}$ and $h_{MF}^{E}$ denotes the mean-field
Hamiltonians acting respectively on the system and the environment. The
mean-field dynamics differs from the exact dynamics since part of the
coupling is not accounted for. Using mean-field expressions, the complete
Hamiltonian can be recast as 
\begin{equation}
H_{I}\left\vert \Psi \right\rangle =\left\{ h_{MF}^{S}\otimes
1_{E}+1_{S}\otimes h_{MF}^{E}-\left\langle A_{\alpha }\right\rangle _{\Phi
}\left\langle B_{\alpha }\right\rangle _{\chi }+\sum_{\alpha }\left(
A_{\alpha }-\left\langle A_{\alpha }\right\rangle _{\Phi }\right) \otimes
\left( B_{\alpha }-\left\langle B_{\alpha }\right\rangle _{\chi }\right)
\right\} \left\vert \Psi \right\rangle   \label{eq:exactmf}
\end{equation}%
In this expression $1_{S}$ and $1_{E}$ correspond to the unity operators
acting respectively on the system and on the environment spaces. Similarly
to the case presented previously, the last term in eq. (\ref{eq:exactmf})
can be reinterpreted as an average over stochastic paths, leading to a new
set of stochastic evolution:

\begin{equation}
\left\{ 
\begin{array}{ll}
d\left\vert \Phi \right\rangle = & \left\{ \frac{dt}{i\hbar }\left(
h_{MF}^{S}-\frac{1}{2}\sum_{\alpha }\left\langle A_{\alpha }\right\rangle _{\Phi
}\left\langle B_{\alpha }\right\rangle _{\chi }\right) +\gamma \sum_{\alpha
}a_{\alpha }\left( A_{\alpha }-\left\langle A_{\alpha }\right\rangle _{\Phi
}\right) \right\} \left\vert \Phi _{\nu }\right\rangle  \\ 
d\left\vert \chi \right\rangle = & \left\{ \frac{dt}{i\hbar }\left(
h_{MF}^{E}-\frac{1}{2}\sum_{\alpha }\left\langle A_{\alpha }\right\rangle _{\Phi
}\left\langle B_{\alpha }\right\rangle _{\chi }\right) +\gamma \sum_{\alpha
}b_{\alpha }\left( B_{\alpha }-\left\langle B_{\alpha }\right\rangle _{\chi
}\right) \right\} \left\vert \chi \right\rangle .
\end{array}%
\right. 
\end{equation}%
Thus the introduction of the mean-field prior to a stochastic formulation
induces a modification of the operator entering the stochastic
contribution. In the latter case $A_{\alpha }$ and $B_{\alpha }$ are
replaced by $A_{\alpha }^{\prime }=(A_{\alpha }-\left\langle A_{\alpha
}\right\rangle _{\Phi })$ and $B_{\alpha }^{\prime }=(B_{\alpha
}-\left\langle B_{\alpha }\right\rangle _{\chi })$. 
{Similar equations have been obtained in ref. \cite{Sha04} using the Hubbard-Stratonovich 
transformation in conjunction with the Girsanov transformation.
More generally when mean-field is introduced prior to 
stochastic formulation, stochastic terms 
generally found in the treatment of Markovian \cite{Gis84,Dio86,Dio94,Ghi90,Bre95} 
as well as non-Markovian dynamics \cite{Dio98,Str99} appear naturally.      
Such a similitude is not surprising since for instance mean-field naturally gives rise 
to orthogonal jumps.
} 
When mean-field is introduced, the
evolution of the statistical fluctuations (noted $\lambda _{stat}^{MF}$),
associated to the stochastic mean-field dynamics with an initial state $%
\left\vert \Psi \right\rangle $, differs from eq. (\ref{eq:norm1}) and reads: 
\begin{equation}
d\lambda _{stat}^{MF}=\frac{dt}{\hbar }\sum_{\alpha }\overline{\left\vert
a_{\alpha }\right\vert ^{2}}\left( \left\langle A_{\alpha }^{+}A_{\alpha
}\right\rangle _{\Phi }-\left\langle A_{\alpha }^{+}\right\rangle _{\Phi
}\left\langle A_{\alpha }\right\rangle _{\Phi }\right) +\overline{\left\vert
b_{\alpha }\right\vert ^{2}}\left( \left\langle B_{\alpha }^{+}B_{\alpha
}\right\rangle _{\chi }-\left\langle B_{\alpha }^{+}\right\rangle _{\chi
}\left\langle B_{\alpha }\right\rangle _{\chi }\right).
\end{equation}%
It is first observed  that the quantum fluctuations for operators $%
A_{\alpha }$ and $B_{\alpha }$ with respect to the state vectors $\left\vert
\Phi \right\rangle $ and $\left\vert \chi \right\rangle $ appear naturally
in this expression. The main advantage of mean-field \cite{Car01,Jul02} 
is that the latter expression is always much smaller than the first
term of eq. (\ref{eq:norm1}) leading generally to smaller statistical
fluctuations.

Let us now combine the advantage of the mean-field with the optimization
proposed in this work. At variance with the general case presented
previously, we have $\left\langle A_{\alpha }^{\prime }\right\rangle _{\Phi
}=\left\langle B_{\alpha }^{\prime }\right\rangle _{\chi }=0$, leading to $%
\Gamma _{\alpha }=0$. As a consequence, only an optimization through the
scaling factor can be performed. In the stochastic mean-field case, the
optimal value is obtained simply by replacing respectively $A_{\alpha }$ and 
$B_{\alpha }$ by $A_{\alpha }^{\prime }$ and $B_{\alpha }^{\prime }$ in
expression (\ref{eq:sfact}). This leads to a reduced statistical fluctuation
given by 
\begin{eqnarray}
\overline{d\left\langle \Psi \left\vert {}\right. \Psi \right\rangle } &=&2%
\frac{dt}{\hbar }\sum_{\alpha }\sqrt{\left( \left\langle A_{\alpha
}^{+}A_{\alpha }\right\rangle _{\Phi }-\left\langle A_{\alpha
}^{+}\right\rangle _{\Phi }\left\langle A_{\alpha }\right\rangle _{\Phi
}\right) \left( \left\langle B_{\alpha }^{+}B_{\alpha }\right\rangle _{\chi
}-\left\langle B_{\alpha }^{+}\right\rangle _{\chi }\left\langle B_{\alpha
}\right\rangle _{\chi }\right) } \\
&\equiv &2\frac{dt}{\hbar }\sum_{\alpha }F_{\alpha }^{MF} . \label{dff_optmf}
\end{eqnarray}%
Using now the fact that 
\begin{eqnarray}
\left( F_{\alpha }\right) ^{2}-\left( F_{\alpha }^{MF}\right) ^{2}=\left( 
\sqrt{\left\langle A_{\alpha }^{+}A_{\alpha }\right\rangle _{\Phi
}\left\langle B_{\alpha }^{+}\right\rangle _{\chi }\left\langle B_{\alpha
}\right\rangle _{\chi }}-\sqrt{\left\langle B_{\alpha }^{+}B_{\alpha
}\right\rangle _{\chi }\left\langle A_{\alpha }^{+}\right\rangle _{\Phi
}\left\langle A_{\alpha }\right\rangle _{\Phi }}\right) ^{2}~~\geq ~~0 ,
\end{eqnarray}%
we see that an optimized stochastic mean-field dynamics will always
further reduce statistical fluctuations.

\subsection{Alternative optimization of the phase factor}

In the stochastic mean-field dynamics presented above, there still exists a
freedom on the phase factor to reduce the statistical errors. More generally
although the method proposed in chapter \ref{secopt} appears as the natural
way to obtain an optimal phase factor, it can obviously not be used if $%
\Gamma _{\alpha }=0$. In this section we propose an alternative strategy to
obtain the phase factor. We assume that $%
\Gamma _{\alpha }$ cancels out. The optimized statistical fluctuations
reduce to 
\begin{eqnarray}
\overline{d\left\langle \Psi \left\vert {}\right. \Psi \right\rangle} =2\frac{dt}{\hbar 
}\sum_{\alpha }\sqrt{\left\langle A_{\alpha }^{+}A_{\alpha }\right\rangle
_{\Phi }\left\langle B_{\alpha }^{+}B_{\alpha }\right\rangle _{\chi }} .
\label{dff_minimal}
\end{eqnarray}%
Here, the SSE without or with mean-field are considered indifferently. In the
latter case, $A_{\alpha }$ and $B_{\alpha }$ must be replaced by $A_{\alpha
}^{\prime }$ and $B_{\alpha }^{\prime }$ in the following expressions. For a
single time step a modification of the phase does not affect directly the
expression (\ref{dff_minimal}). However the phase factor can be adjusted to
act on the value of $c_{2}^{\alpha }=\left\langle A_{\alpha }^{+}A_{\alpha
}\right\rangle _{\Phi }\left\langle B_{\alpha }^{+}B_{\alpha }\right\rangle
_{\chi }$ along the trajectory. Indeed, starting from the SSE given by equation (\ref{eq:sse}), we
have \footnote{
in the mean-field case, additional terms exist due to the deterministic
part in the SSE. These terms are not reported here but will not change the
conclusion.} 
\begin{eqnarray}
\begin{array}{lll}
\overline{dc_{2}^{\alpha }} & = & ~~\frac{dt}{\hbar }\sum_{\beta
}\left\langle A_{\beta }^{+}A_{\alpha }^{+}A_{\alpha }A_{\beta
}\right\rangle _{\Phi }\left\langle B_{\alpha }^{+}B_{\alpha }\right\rangle
_{\chi } \\ 
&  & +\frac{dt}{\hbar }\sum_{\beta }\left\langle A_{\alpha }^{+}A_{\alpha
}\right\rangle _{\Phi }\left\langle B_{\beta }^{+}B_{\alpha }^{+}B_{\alpha
}B_{\beta }\right\rangle _{\chi } \\ 
&  & +\frac{dt}{i\hbar }\sum_{\beta }\overline{a_{\beta }b_{\beta
}}\left\langle A_{\alpha }^{+}A_{\alpha }A_{\beta }\right\rangle _{\Phi
}\left\langle B_{\alpha }^{+}B_{\alpha }B_{\beta }\right\rangle _{\chi } \\ 
&  & -\frac{dt}{i\hbar } \sum_{\beta } \overline{a_{\beta }^{\ast
}b_{\beta }^{\ast }}\left\langle A_{\beta }^{+}A_{\alpha }^{+}A_{\alpha
}\right\rangle _{\Phi }\left\langle B_{\beta }^{+}B_{\alpha }^{+}B_{\alpha
}\right\rangle _{\chi } \\ 
&  & +2\frac{dt}{\hbar }\Re \left\{ \sum_{\beta }\overline{a_{\beta
}b_{\beta }^{\ast }}\left\langle A_{\alpha }^{+}A_{\alpha }A_{\beta
}\right\rangle _{\Phi }\left\langle B_{\beta }^{+}B_{\alpha }^{+}B_{\alpha
}\right\rangle _{\chi }\right\} 
\end{array}%
\end{eqnarray}%
In this expression, only the last term is influenced by a phase-factor. The
others are invariant with respect to a phase transformation. In order to
reduce the statistical fluctuations, it is suitable to drive the system
towards small values of $\overline{c_{2}^{\alpha }}$ during the evolution. This can be
achieved by means of a proper adjustment of $\theta _{\alpha }$. Similarly
to section \ref{optph}, we define: 
\begin{eqnarray}
\left\langle A_{\alpha }^{+}A_{\alpha }A_{\beta }\right\rangle _{\Phi
}\left\langle B_{\beta }^{+}B_{\alpha }^{+}B_{\alpha }\right\rangle _{\chi
}=e^{i\theta _{AB}^{\prime }}\left\vert \left\langle A_{\alpha
}^{+}A_{\alpha }A_{\beta }\right\rangle _{\Phi }\left\langle B_{\beta
}^{+}B_{\alpha }^{+}B_{\alpha }\right\rangle _{\chi }\right\vert 
\end{eqnarray}%
and the new optimal phase factor is then given by $2\theta _{\alpha
}=\pi -\theta _{AB}^{\prime }$ leading to 
\begin{eqnarray}
\Re \left( \overline{a_{\beta }b_{\beta }^{\ast }}\left\langle A_{\alpha
}^{+}A_{\alpha }A_{\beta }\right\rangle _{\Phi }\left\langle B_{\beta
}^{+}B_{\alpha }^{+}B_{\alpha }\right\rangle _{\chi }\right) =-\left\vert
\left\langle A_{\alpha }^{+}A_{\alpha }A_{\beta }\right\rangle _{\Phi
}\left\langle B_{\beta }^{+}B_{\alpha }^{+}B_{\alpha }\right\rangle _{\chi
}\right\vert 
\end{eqnarray}%
As previously we assume that $\left\vert \overline{a_{\beta }b_{\beta
}^{\ast }}\right\vert =1$. The method presented here provides an indirect
way to reduce statistical fluctuations when the second term of equation (\ref%
{eq:norm1}) is of  no use. This is always the case in
stochastic mean-field dynamics. Therefore in the forthcoming applications
of stochastic mean-field, the phase will always be deduced from the latter alternative method.

\section{Illustration}

In this chapter we apply the different optimization procedures
described previously on an illustrative example consisting of a system of
spins in interaction. We show in this example that the statistical
fluctuations can be significantly reduced leading to more efficient
stochastic calculations. A systematic study shows that the reduction
of statistical fluctuations depends on the parameters of the model.

The analytical
model proposed in  \cite{Bre04} is considered. The model considers a spin described by its
Pauli operator $\sigma $ coupled with an environment of spins with spin
operators $\sigma ^{(\alpha )}$ ($\alpha =1,\cdots, N$) where $N$ is the
number of spins. The interaction Hamiltonian is given by 
\begin{eqnarray}
H=2 \sum_{\alpha }C_{\alpha }\left( \sigma _{+}\sigma _{-}^{(\alpha )}+\sigma
_{-}\sigma _{+}^{(\alpha )}\right) 
\end{eqnarray}%
where $C_{\alpha }$ is the coupling constant. This model is a simplified
version of the one used for the description of a single electron spin in a
quantum dot given in  \cite{Kha02}. Its simplicity is particularly suitable
for the present study, enabling us to focus mainly on statistical
fluctuations. In addition it has a similar form to the one generally taken 
in the description of open quantum systems\cite{Gar00,Coh97}. Here, it is assumed
that $C_{\alpha }=C/N$. In this case starting from an initial density $\rho
\left( t=0\right) =\left\vert \Psi \left( 0\right) \right\rangle
\left\langle \Psi \left( 0\right) \right\vert $ with 
\begin{eqnarray}
\left\vert \Psi \left( 0\right) \right\rangle =\left\vert +\right\rangle
\otimes \left\vert -,\cdots ,-\right\rangle, 
\end{eqnarray}%
the dynamical evolution is known analytically \cite{Bre04}. In particular
the exact system density evolution obtained by performing the partial trace
on the environment, i.e. $\rho _{S}\left( t\right) =Tr_{E}\left( \rho \left(
t\right) \right) $, is given by 
\begin{eqnarray}
\rho _{S}\left( t\right) =\rho _{++}\left\vert +\right\rangle \left\langle
+\right\vert +\rho _{--}\left\vert -\right\rangle \left\langle -\right\vert 
\end{eqnarray}%
with $\rho _{++}=1-\rho _{--}=\cos ^{2}\left( 2Ct\right) $. We consider the
model as a benchmark for the different SSE considered in this work. In the
present case the interaction operators discussed in the previous chapters
read: $A_{\alpha }=\sqrt{\left( \frac{2C}{N}\right) }\sigma _{\pm }$ and $%
B_{\alpha }=\sqrt{\left( \frac{2C}{N}\right) }\sigma _{\pm }^{(\alpha )}$.

We now consider four different classes of simulations. In the first type of
calculations mean-field is not accounted for. Two calculations will be
performed without (noted SSE) and with optimization 
(OSSE). The SSE case can  be
considered as the reference case and corresponds to $a_{\alpha }=b_{\alpha
}=x_{\alpha }$. In the second type of calculations the mean-field is
introduced and the cases without and with optimization are again considered.
They will be referred to as Stochastic Mean-field (SMF), again with $a_{\alpha
}=b_{\alpha }=x_{\alpha }$ and Optimized Stochastic Mean-Field (OSMF).

\subsection{Statistical fluctuations}

\begin{figure}[tbph]
\includegraphics[height=10.cm,angle=-90.]{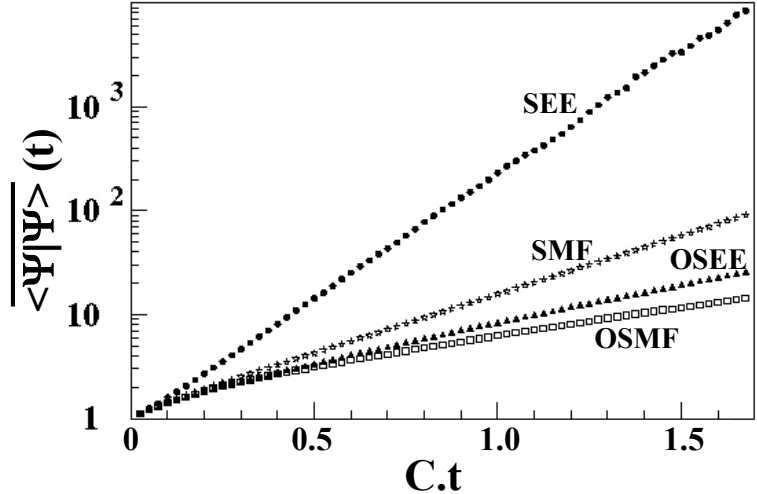}
\caption{Evolution of $\overline{\left\langle \Psi \left\vert {}\right. \Psi
\right\rangle }$ (in log scale) as a function of time $t$ multiplied by the coupling constant 
$C$ for $N=1$ and $C=0.5$ (in arbitrary units). The calculations correspond to SSE (filled
circles), OSSE (filled triangles), SMF (open stars) and OSMF (open squares).
Averages have been obtained with $10^{5}$ trajectories.}
\label{fig:stat1}
\end{figure}

Let us first consider  the case $N=1$ and $C=0.5$. In order to illustrate
quantitatively the effect of the different optimization, we perform an
ensemble of stochastic trajectories all starting from the same initial state
vector $\left\vert \Psi (0)\right\rangle $. 
The evolution in time of the average quantity $\overline{\left\langle \Psi
\left\vert {}\right. \Psi \right\rangle }$ is shown in figure \ref{fig:stat1}. 
The SSE, OSSE, SMF and OSMF
calculations are represented respectively by filled circles, filled
triangles, open stars and open squares. In each case an ensemble of $%
N_{traj}=10^{5}$ trajectories are performed. Let us first focus on the
stochastic dynamics without optimization. We see that $\overline{%
\left\langle \Psi \left\vert {}\right. \Psi \right\rangle }$ increases much
faster in the standard SSE case than in Stochastic mean-field dynamics. As
expected the introduction of the mean-field alone reduces
statistical fluctuations. When optimization procedures are introduced, the
evolution of $\overline{\left\langle \Psi \left\vert {}\right. \Psi
\right\rangle }$ is significantly further reduced. It is worth  noticing  that
the OSSE gives a much lower increase than the SMF dynamics. This indicates
that the effect of optimization on the standard SSE already gives a much
better result than the one using stochastic mean-field without optimization.
Finally Fig. \ref{fig:stat1} shows that the maximal reduction is obtained
when both the mean-field and the optimization are taken into account.

We would like to mention that the present model has several specific
aspects. First if the phase factor is not optimized (i.e. the phase is set
to zero), no improvement is observed. Indeed if the original noise is set to 
$a_{\alpha }=b_{\alpha }=x_{\alpha }$, then $\left\langle A_{\alpha
}^{+}A_{\alpha }\right\rangle _{\Phi }=\left\langle B_{\alpha }^{+}B_{\alpha
}\right\rangle _{\chi }$ along the trajectory, leading to an optimal scaling
factor constantly equal to one. Therefore the strong reduction is due to
the combined action of the phase factor and the scaling factor. Note that in
the present model, it is possible to show that for a given initial state $%
\theta _{AB}=\theta _{AB}^{\prime }$,  the optimal phases are the same in
the two optimized calculations.

\begin{figure}[tbp]
\begin{center}
\includegraphics[height=10.cm,angle=-90.]{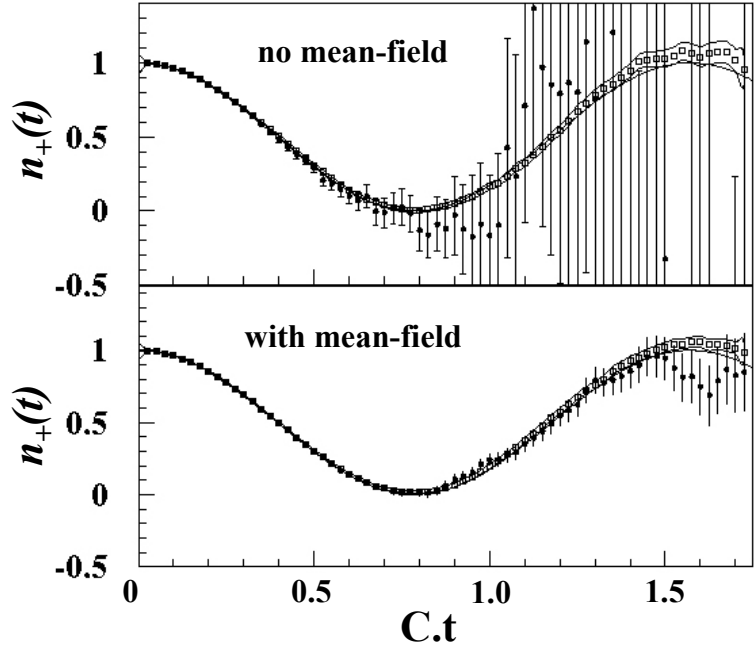}
\end{center}
\caption{The occupation probability $n_+$ obtained by averaging over
stochastic trajectories is shown as a function of as a function of time $t$ 
multiplied by the coupling constant $C$ and compared to the
exact dynamics (solid line) with $10^4$ trajectories. Top: Results obtained
using the SSE with (open squares) and without (filled circles) optimization.
Bottom, results obtained using the SMF techniques with (open squares) and without
(filled circles) optimization. In each case the error bars represent the
standard deviations.}
\label{fig:stat2}
\end{figure}

\begin{figure}[tbp]
\begin{center}
\includegraphics[height=10.cm,angle=-90.]{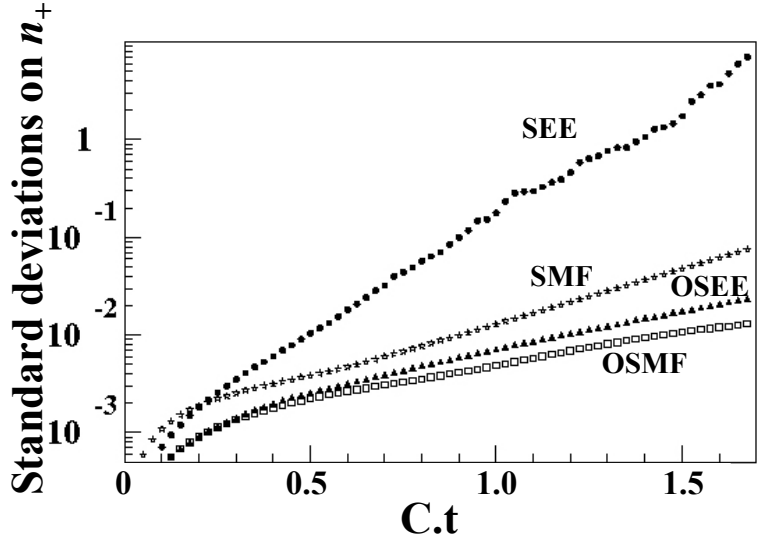}
\end{center}
\caption{Evolution of the standard deviations of $n_+$ 
as a function of time $t$ multiplied by the coupling constant 
$C$ obtained for $10^5$
trajectories using SSE (filled circles), OSSE (filled triangles), SMF (open
stars) and OSMF (open squares).}
\label{fig:stat3}
\end{figure}

In order to quantify the statistical fluctuation growth, each curve is
fitted using the function $\overline{ \left\langle \Psi \left. {}\right\vert \Psi
\right\rangle} \propto e^{\lambda _{s}t}$. A measurement of the increase is
directly given by $\lambda _{s}$. We have obtained $\lambda _{s}\simeq 2.6$, 
$1.3$, $0.78$ and $0.53$  for the SSE, SMF, OSSE and OSMF cases respectively.
The reduction of the statistical errors has a direct effect on the capacity
of the different stochastic formulations to efficiently reproduce the exact
dynamics: the slower the increase the more efficient the stochastic
equation. This is illustrated in figure \ref{fig:stat2} where the evolution
of the occupation probability $n_{+}(t)= \left<+\left|\rho_S(t)
\right|+\right>$ 
is shown as a function of time for the four calculations and
compared to the exact result. In all cases the same number of
trajectories $N_{traj}=10^{4}$ was used. In each case the standard deviation
on $n_{+}$ calculated with the set of trajectories is shown in the form of 
error bars. Large values are observed in the case of standard SSE. In
addition the SSE  calculation is unable to reproduce the exact dynamics
over the considered time interval. The optimized calculations give a good
reproduction of the exact dynamics while in the SMF a small departure
from the exact solution is observed at large time. A precise comparison
indicates (as expected) that the OSMF gives the best agreement. These
results show again that the reduction of the statistical fluctuations
improves the stochastic methods. For a given number of trajectories, the
associated standard deviations are proportionally reduced. This is
illustrated in figure \ref{fig:stat3} where the standard deviations of $n_{+}$
obtained with the four types of calculations are displayed. 
Standard deviations follow the same trends as the statistical
fluctuations displayed in figure \ref{fig:stat1}. Note that all stochastic
methods lead to the exact result for an infinite number of trajectories .
However in order to have the same standard deviations with the standard SSE as
the one displayed in figure \ref{fig:stat2} using OSMF, several millions
of trajectories are needed.


\subsection{Variation of parameters}

\begin{figure}[tbp]
\begin{center}
\includegraphics[height=10.cm,angle=-90.]{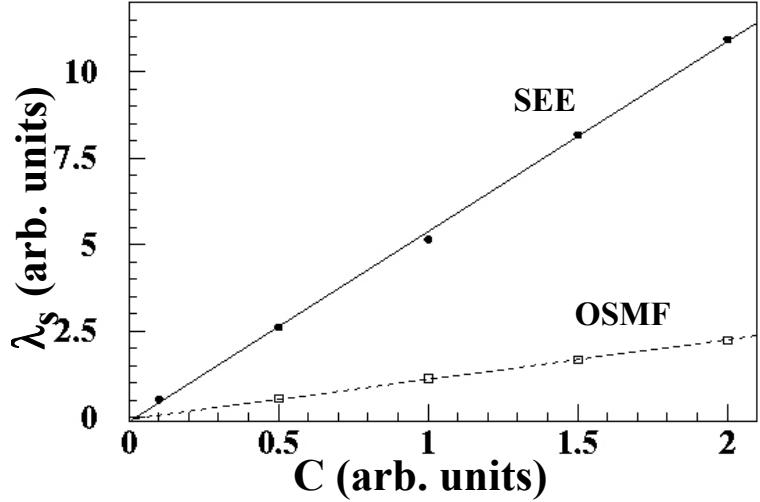}
\end{center}
\caption{ Evolution of $\protect\lambda_s$ as a function of the coupling
constant $C$ obtained using SSE (filled circles) and OSMF (open squares).
The result of a fit is displayed with solid (for SSE) and dashed (for OSMF) lines 
and corresponds respectively to $\protect\lambda _s\propto 5.5 C$ and $%
\protect\lambda _s \propto 1.15 C$.}
\label{fig:stat4}
\end{figure}

\begin{figure}[tbp]
\begin{center}
\includegraphics[height=10.cm,angle=-90.]{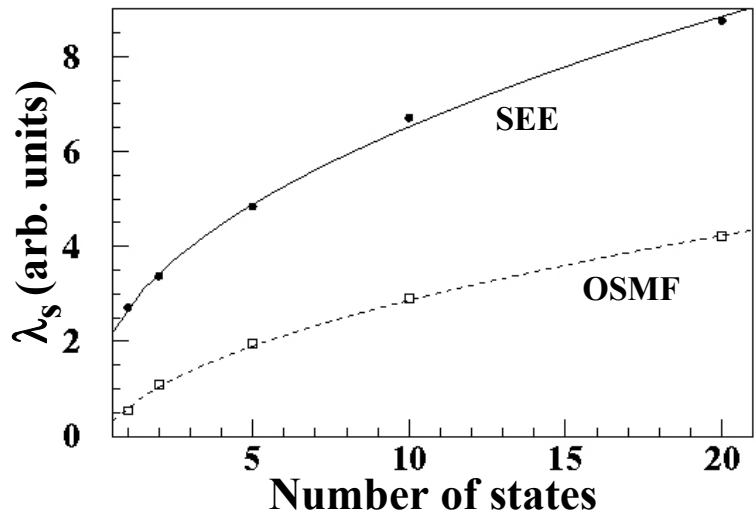}
\end{center}
\caption{ Evolution of $\protect\lambda_s$ as a function of the number of
states $N$ in the bath for $C= 0.5$ obtained using SSE (filled circles) and
OSMF (open squares). The result of a fit is displayed with solid and dashed
lines for the SSE and OSMF cases and corresponds to $\protect\lambda %
_s \propto 1.8 \protect\sqrt{N}$ and $\protect\lambda _s \propto 1.0 \protect%
\sqrt{N}$.}
\label{fig:stat5}
\end{figure}

We now systematically study the effect of the different
optimizations on the behavior of statistical fluctuations when the parameters of the model
change. In a first series of calculations, the effect of
the variation of the coupling constant $C$ for a fixed number of bath states (%
$N=1$) is considered. In the second series of calculations the number of states was varied while $%
C=0.5$. $\lambda _{s}$ was obtained by fitting the statistical
fluctuations. We show in figure \ref{fig:stat4}
and figure \ref{fig:stat5} the values of $\lambda _{s}$ respectively as
a function of $C$ and $N$ for the SSE and the OSMF calculations. In figure \ref{fig:stat4} for both
stochastic calculations, $\lambda _{s}$ exhibits a linear behavior with $C$ in agreement 
with the expressions (\ref{eq:norm1})
and (\ref{dff_optmf}). However the statistical fluctuations
increase much more rapidly with $C$ in the SSE case than in the OSMF case.
Therefore it is found that the effect of the optimization increases
when the coupling increases. Similarly $\lambda _{s}$ increases faster with
the number $N$ of states in the bath in SSE compared to the optimized cases.
Again the optimization is more efficient when $N$ increases.

In this section we have illustrated by a model case the optimization
procedure described previously. We have shown that the combined
effect of using the mean-field and the adaptative noise significantly improves
the quantum jump simulations. Lastly, we have shown that the improvement depends
explicitly on the coupling strength as well as on the number of states in
the bath.


\section{Conclusion}

In this work the freedom in the stochastic reformulation of the exact
dynamics of interacting systems using pairs of wave-packets 
has been used to optimize stochastic trajectories.
For a given initial state an optimal complex
noise that minimizes statistical fluctuations is obtained. It is shown that
the properties of the stochastic variables depend on the specific state to
which it is applied as well as on the interaction properties, hence 
the denomination '\textit{adaptative noise method}'. We have also shown
that the method can be combined with the introduction of a 
mean-field prior to the stochastic reformulation. The combination of the mean-field
and the adaptative noise is expected to lead in general to a further reduction
of the statistical fluctuations. 

The different optimization procedures have been studied with a schematic model consisting 
of a single spin interacting with a bath of spins. In this model the
optimization of jumps significantly reduces the statistical fluctuations.
Accordingly it has been shown that for a given number of trajectories, a much better
reproduction of the exact dynamics is obtained compared to the standard SSE
without optimization. Although promising, more complex 
models should be used to  fully demonstrate the power of the method.

At last we would like to mention that the work described here cannot be
directly applied to stochastic Many-Body theories to treat self-interacting bosonic or fermionic systems 
as proposed in ref. \cite{Car01,Jul02}.
In this case, the operators $A_\alpha$ and $B_\alpha$ are identical and
are applied to the same state. Work is currently in progress to modify the
original stochastic reformulation in order to take advantage of the
optimization procedure described here.

{\bf ACKNOWLEDGMENTS}

The author is grateful to Jean-Luc Lecouey, Dominique Durand, Daniel Cussol and John Frankland for 
the careful reading of the manuscript.

\end{document}